\documentclass[conference]{IEEEtran}
\IEEEoverridecommandlockouts

\usepackage{enumitem}
\usepackage{booktabs}
\usepackage{multirow}
\usepackage{flushend}
\usepackage{graphicx}
\usepackage{amsmath}
\usepackage{adjustbox}
\usepackage[colorlinks=true]{hyperref}

\begin{document}
    \title{Can No-Reference Quality-Assessment Methods\\Serve as Perceptual Losses for Super-Resolution?\\
    \thanks{\IEEEauthorrefmark{1}These authors contributed equally to this work}
}

\author{
    \IEEEauthorblockN{
        Egor Kashkarov\IEEEauthorrefmark{1}
        \quad
        Egor Chistov\IEEEauthorrefmark{1}
        \quad
        Ivan Molodetskikh
        \quad
        Dmitriy Vatolin
    }
    \IEEEauthorblockA{
        \textit{Lomonosov Moscow State University}
    }
    \IEEEauthorblockA{
        \texttt{\{egor.kashkarov, egor.chistov, ivan.molodetskikh, dmitriy\}@graphics.cs.msu.ru}
    }
}

\maketitle

\begin{abstract}
Perceptual losses play an important role in constructing deep-neural-network-based methods by increasing the naturalness and realism of processed images and videos. Use of perceptual losses is often limited to LPIPS, a full-reference method. Even though deep no-reference image-quality-assessment methods are excellent at predicting human judgment, little research has examined their incorporation in loss functions. This paper investigates direct optimization of several video-super-resolution models using no-reference image-quality-assessment methods as perceptual losses. Our experimental results show that straightforward optimization of these methods produce artifacts, but a special training procedure can mitigate them.
\end{abstract}


    \section{Introduction}
Deep perceptual loss is a loss function that aims to reproduce human perception by way of deep features obtained from neural networks. These perceptual losses are important to improving content’s naturalness and realism. Since the pioneering work of Johnson et al.~\cite{johnson2016perceptual}, video-restoration methods have employed these losses heavily to achieve results that are more visually plausible. But use of perceptual losses is often limited to LPIPS~\cite{zhang2018unreasonable}.

Recent work demonstrates that video super-resolution (VSR) is generalizable to other video-restoration tasks, such as denoising and deblurring~\cite{chan2022generalization}. Our study focuses on VSR, but our conclusions apply to other fields as well.

New neural-network-based approaches to no-reference image-quality-assessment (IQA), such as MDTVSFA~\cite{li2021unified} and CLIP-IQA~\cite{wang2023exploring}, are excellent at predicting human judgment. Tools such as IQA-PyTorch allow easy optimization of IQA methods. Still unclear, however, is whether optimization of these methods can improve visual quality.

\begin{figure}[t]
    \includegraphics[width=\columnwidth]{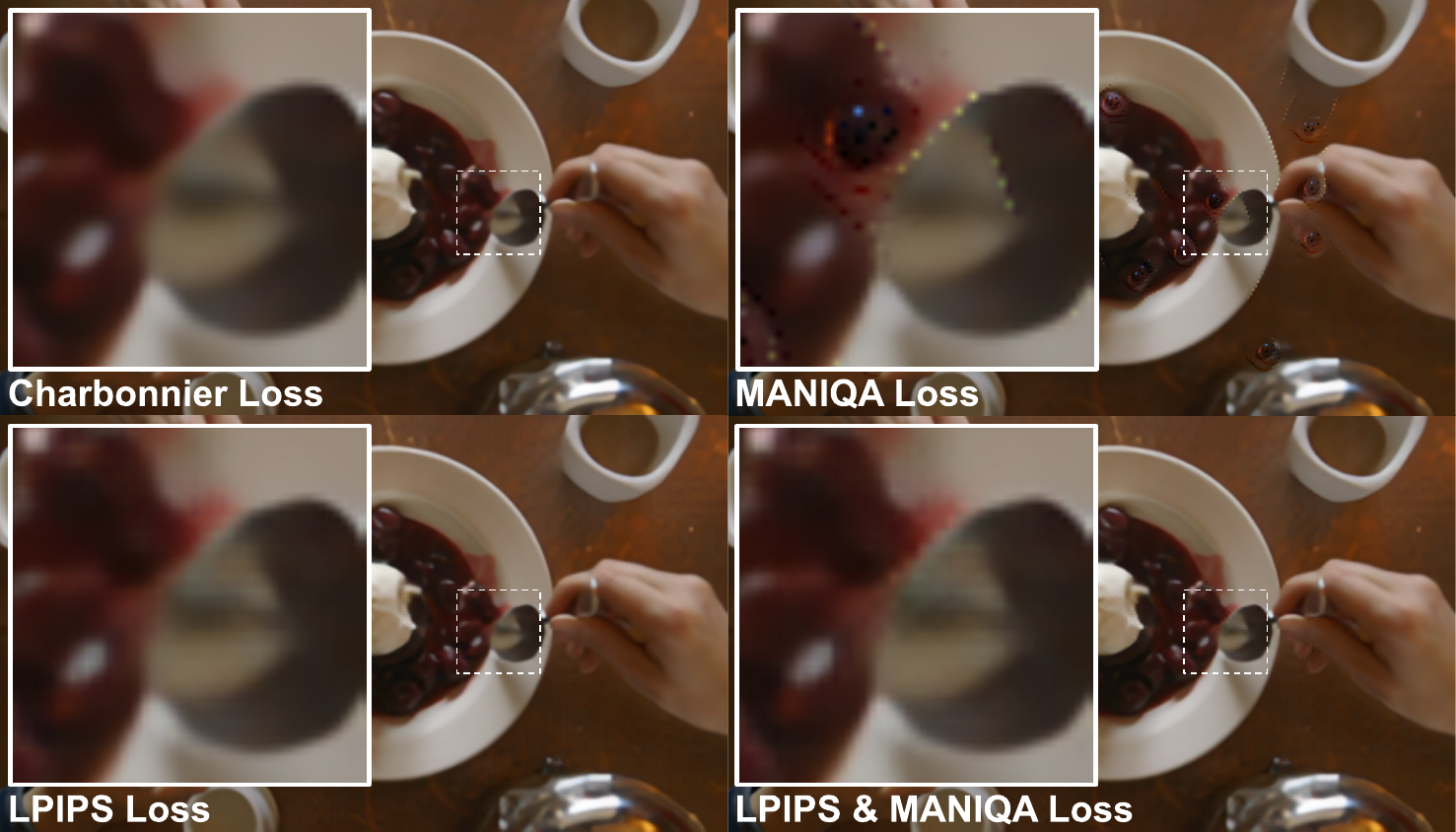}
    \caption{BasicVSR++ on a test sequence from Vimeo-90K. Fine-tuning with MANIQA produces green dotted edges and "eyes," while fine-tuning using LPIPS combined with MANIQA does not. Although the bottom row exhibits a barely noticeable difference, optimizing the two losses yields a higher average gain over the 11 IQA methods we evaluated.}
    \label{fig:example}
\end{figure}

Optimizing just perceptual loss is sometimes insufficient to accurately recover local and diverse shapes in an image, potentially leading to unwanted artifacts and artificial patterns. Fig.~\ref{fig:example} shows an example. Loss functions can be combined, but finding the optimal composition is still difficult.

\begin{table*}[t]
    \caption{Weights for each loss function in our experiments. Increasing the weights often led to gradient explosion. FR stands for full-reference IQA methods, and NR stands for no-reference IQA methods.}
    \label{tab:weights}
    \centering
    \renewcommand{\arraystretch}{0.85} 
    \begin{tabular}{lcccccccccc}
\toprule
\multirow{2}{*}{Loss Function} & Charbonnier & LPIPS & PieAPP & DBCNN & NIMA & PaQ-2-PiQ & HyperIQA & MDTVSFA & MANIQA & CLIP-IQA \\
 & 1994 & 2018 & 2018 & 2018 & 2018 & 2020 & 2020 & 2021 & 2022 & 2023 \\
\midrule
Type & FR & FR & FR & NR & NR & NR & NR & NR & NR & NR \\
Weight & +1 & +0.05 & +0.00005 & -0.005 & -0.005 & -0.00005 & -0.005 & -0.005 & -0.005 & -0.005 \\
\bottomrule
\end{tabular}

\end{table*}
 
Researchers have studied the behavior of neural-network-based approaches when fine-tuning with different IQA methods. For instance, Ding et al~\cite{ding2021comparison}. looked at four image-restoration tasks and 11 full-reference IQA methods; they showed that deep perceptual losses such as LPIPS and DISTS~\cite{ding2020image} outperform classical ones in three out of four applications. Mohammadi et al.~\cite{mohammadi2022perceptual} confirmed these conclusions on a deep-image-coding task.

Pihlgren et al.~\cite{pihlgren2023systematic} studied four other tasks and seven loss networks based on deep-image-classification architectures. They suggested using the VGG architecture (on which LPIPS and DISTS are based) for optimization and found that selecting the feature-extraction point is at least as important as selecting the architecture.

Differing from previous approaches that seek to optimize full-reference IQA methods, our contributions are the following:

\begin{itemize}[itemsep=\topsep]
    \item We demonstrate how optimization with no-reference IQA methods affects the scores of other methods as well as the visual artifacts that optimization can produce;
    \item We demonstrate that a specific training procedure can mitigate these artifacts by optimizing a combination of quality-assessment methods.
\end{itemize}

\section{Methods}
This work focuses on fine-tuning video super-resolution (VSR) models using deep no-reference image-quality-assessment (IQA) methods as perceptual losses.

\subsection{Choosing VSR and IQA Methods}
We studied three architectures and approaches to training neural networks: recurrent neural networks (RNNs), generative adversarial networks (GANs), and transformers. This strategy allowed us to cover most of the model types that have served in video restoration and other image-to-image tasks over the past few years.

For each architecture, we selected one VSR method. We chose BasicVSR++~\cite{chan2022basicvsr++}, iSeeBetter~\cite{chadha2020iseebetter}, and VRT~\cite{liang2024vrt}, respectively, for the RNNs, GANs, and transformers. These methods show good performance on multiple VSR benchmarks~\cite{vsrpwc}.

Because previous work focused primarily on optimizing classical full-reference IQA methods such as MSE, SSIM~\cite{wang2004image}, GMSD~\cite{xue2013gradient}, and VSI~\cite{zhang2014vsi} and on comparing them with LPIPS~\cite{zhang2018unreasonable}, we decided to instead consider no-reference differentiable methods. Our full list includes DBCNN~\cite{zhang2018blind}, NIMA~\cite{talebi2018nima}, PaQ-2-PiQ~\cite{ying2020patches}, HyperIQA~\cite{su2020blindly}, MDTVSFA~\cite{li2021unified}, MANIQA~\cite{yang2022maniqa}, and CLIP-IQA~\cite{wang2023exploring}. We also study two full-reference methods: LPIPS~\cite{zhang2018unreasonable} and PieAPP~\cite{prashnani2018pieapp}.

\subsection{IQA-Method Weights in Loss Function}
In all our experiments we trained a combination of the Charbonnier loss function, which the BasicVSR++ developers used, and another loss function based on some of the IQA methods. Different IQA methods produce different gradient values, so clearly some weighting is necessary. We reweighted all loss components proportionally to their typical values for real images; Table~\ref{tab:weights} shows these values.

We also tried using the GradNorm~\cite{chen2018gradnorm} algorithm to balance losses in our composition. GradNorm has a hyperparameter $\alpha,$ for controlling the restoring force that brings the losses back to a common weight. We chose $\alpha=1.5$, which the authors recommend. In our experiments, GradNorm performed worse than the empirically selected weights. Besides low results, the GradNorm reference implementation often caused optimization errors when using two or more IQA methods.

We found that optimization of some IQA methods can impair the judgment of other studied ones. Hence, we also studied different loss-function combinations to find the one that best improves most of the IQA methods. When fine-tuning IQA-method combinations, we employed the same weights in Table~\ref{tab:weights}.

\section{Experiments}
To compare the performance of our chosen IQA methods as perceptual losses, we trained several VSR models with them and used classical methods such as PSNR and SSIM. We also selected nine deep IQA methods for the comparison.

\subsection{Dataset Selection}
Training employed the Vimeo-90K~\cite{xue2019video} dataset. For evaluation we selected 100 videos from that same dataset and 30 videos from REDS~\cite{nah2019ntire}; we also filmed 35 videos ourselves using 14 Apple, Samsung, Google, Xiaomi, and Huawei smartphones as well as one action camera. Our dataset contains outdoor scenes that exhibit different motion types and lighting as well as different seasons and times of day. All videos underwent downscaling by a factor of four using bicubic interpolation to achieve low-resolution frames. As in Vimeo-90K, we limited all video lengths to seven frames.

\begin{figure}[htbp]
    \includegraphics[width=\columnwidth]{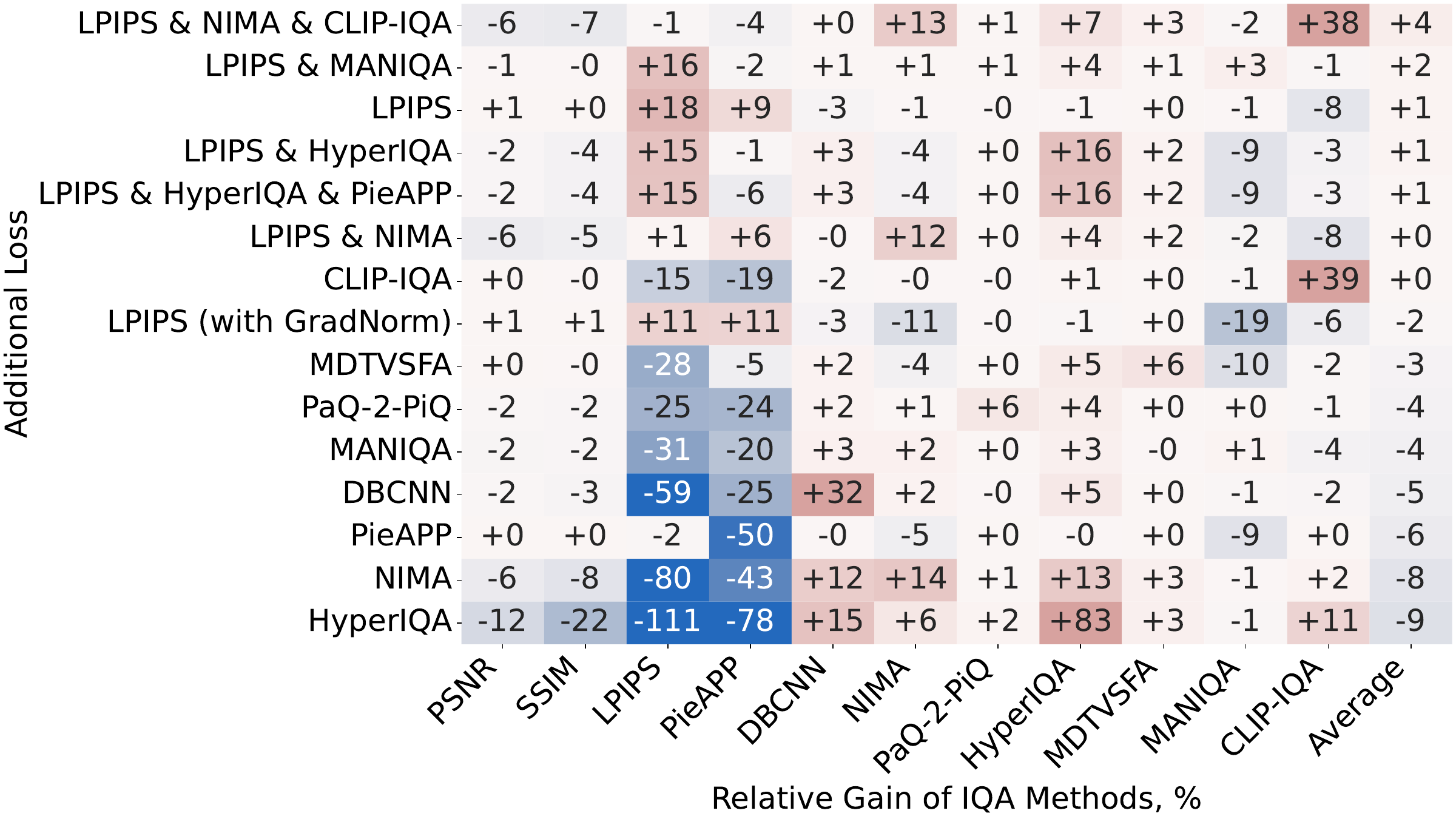}
    \caption{Relative gain for IQA methods after optimization with different IQA-method combinations. Rows represent an additional component to the Charbonnier loss function and columns represent averaged ralative gain over different IQA-metrics. We averaged the results over BasicVSR++, iSeeBetter, VRT, and three datasets. Although LPIPS and PieAPP react badly to the tuning of most no-reference IQA methods, loss-function combinations improve the judgment of nearly all IQA methods.}
    \label{fig:heatmap}
\end{figure}

\begin{figure*}[t]
    \includegraphics[width=\textwidth]{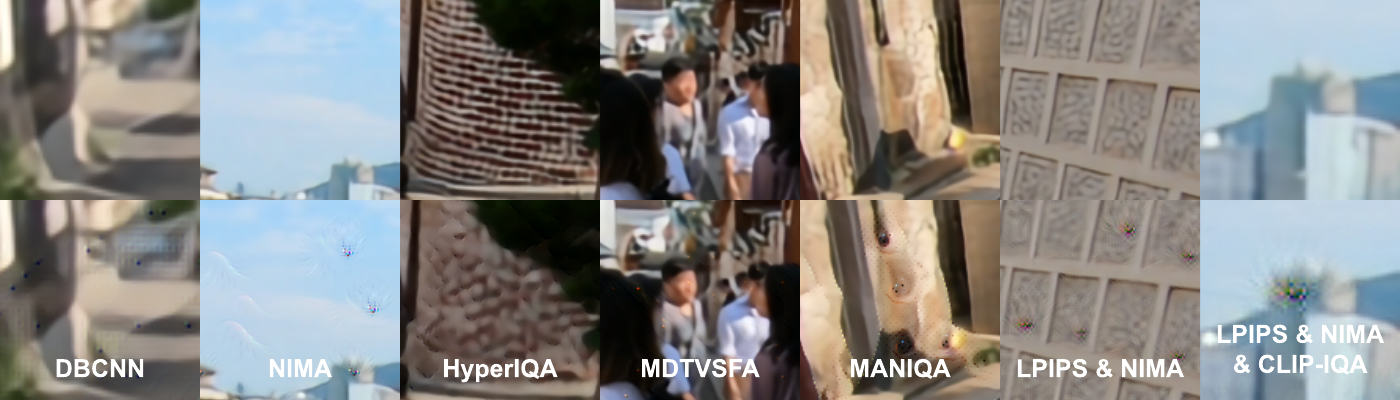}
    \caption{BasicVSR++ on a test sequence from REDS. The top row is a baseline model trained with Charbonnier loss only. Fine-tuning with DBCNN, NIMA, MDTVSFA, and MANIQA produces "colored dot" artifacts, and fine-tuning with HyperIQA blurs the image. Even when used in combinations, NIMA still creates artifacts.}
    \label{fig:artifacts}
\end{figure*}

\subsection{Training Procedure}
We trained baseline VSR models until convergence with the Charbonnier loss function, as the BasicVSR++ developers did. They chose this loss function because it tends to work better than L1 or L2~\cite{lai2017deep}. Training took 5k iterations for BasicVSR++ and 50k iterations for VRT and iSeeBetter. We then fine-tuned the trained models to convergence using a combination of the Charbonnier loss function and various IQA methods. This fine-tuning took 15k more iterations for BasicVSR++ and 25k iterations more for VRT and iSeeBetter.

Both the training and fine-tuning used the Adam optimizer with a $0.0004$ learning rate and a cosine-annealing scheduler with warmup and restarts. For convenience, we chose the SpyNet~\cite{ranjan2017optical} optical-flow algorithm for all VSR models. Note that only BasicVSR++ and VRT employ SpyNet as their optical-flow algorithm in the reference implementations.

\subsection{Evaluation}
To evaluate the contribution of different loss functions, we calculated the IQA-method's values $M$ and averaged them over testing dataset for both baseline model $B,$ trained with only the Charbonnier loss function, and finetuned model $T,$ trained with composite loss function. The relative gain was calculated then as follows: 
\begin{equation}
    \text{Relative Gain} = 100s \frac{M_T - M_B}{M_B},
\end{equation}
where $s = +1$ when higher values of an IQA method are better and $s = -1$ otherwise. The results are shown in the Fig.~\ref{fig:heatmap}. We found nearly identical results among three VSR models and three datasets, so we provide relative gains averaged over both datasets and models. We also added Charbonnier and LPIPS composition with weights tuned using GradNorm to demonstrate its low performance.

\section{Results and Discussion}
We showed that fine-tuning with different losses can lead to diverging situations. Most IQA methods, when serving as loss functions, can improve themselves — sometimes to extreme values, such as 83\% for HyperIQA. But PieAPP is different for reasons that remain unclear, as even small weights cause judgment degradation.

Another observation is that full-reference deep IQA methods such as LPIPS and PieAPP respond poorly to fine-tuning of no-reference IQA methods. The result is judgment deterioration of up to 111\%. We believe the reason is that no-reference IQA methods, when undergoing optimization, often produce low-magnitude noise and artificial patterns in an image. These patterns are similar to those generated by adversarial attacks as described in Antsiferova et al.~\cite{antsiferova2024comparing}.

Nearly all no-reference IQA methods produce some artificial patterns that reduce overall quality. Fig.~\ref{fig:artifacts} shows examples. In our experiments, only PaQ-2-PiQ and CLIP-IQA produced none of these patterns. For PaQ-2-PiQ small weight values apparently prevent overfitting.

In combination with LPIPS, all no-reference IQA methods cease to degrade LPIPS and PieAPP. The artificial patterns also diminish for these loss-function combinations. Only LPIPS combined with NIMA as well as LPIPS combined with NIMA and CLIP-IQA, retain artificial patterns identical to the NIMA ones. We recommend also adding LPIPS to the loss function when fine-tuning VSR models with no-reference IQA methods.

Our final observation is that classical IQA methods, such as PSNR and SSIM, are almost insensitive to fine-tuning with deep IQA methods. Their use is also insufficient for finding all the artificial patterns that optimization can produce. Apparently, however, LPIPS measurements can aid in detecting artificial patterns. Huge LPIPS deterioration indicates the presence of these patterns when that perceptual loss was not optimized directly; small LPIPS changes instead of a large improvement indicate the presence of artificial patterns when it was optimized directly.

\section{Conclusion and Future Work}
In this paper we investigate direct optimization of deep image-quality-assessment methods and their combinations. Our experimental results show how optimizing with different IQA methods affects other IQA methods and can cause visual artifacts. We provide example artifacts and demonstrate that optimizing the combination of LPIPS and a no-reference IQA method can mitigate them. 

The final answer to the question in the title is that the use of no-reference quality-assessment methods as perceptual losses for super-resolution task is not fully justified. Despite the fact that a certain quantitative gain can be achieved, the emerging visual artifacts negate this advantage.

Future work on this topic could include more-extensive experiments with a larger number of IQA methods and their combinations. Additionally, we have planned more-focused efforts to select loss-component weights on the basis of their gradient norm. The goal is to achieve a visual-quality improvement and validate it through subjective comparison.

    \bibliographystyle{IEEEtran}
    \bibliography{IEEEabrv,index}
\end{document}